\documentclass[10pt,twoside,twocolumn,british,english,preprint]{revtex4}
\usepackage[T1]{fontenc}
\usepackage[latin9]{inputenc}
\usepackage{float}
\usepackage{amssymb}
\usepackage{graphicx}

\makeatletter
\@ifundefined{textcolor}{}
{%
 \definecolor{BLACK}{gray}{0}
 \definecolor{WHITE}{gray}{1}
 \definecolor{RED}{rgb}{1,0,0}
 \definecolor{GREEN}{rgb}{0,1,0}
 \definecolor{BLUE}{rgb}{0,0,1}
 \definecolor{CYAN}{cmyk}{1,0,0,0}
 \definecolor{MAGENTA}{cmyk}{0,1,0,0}
 \definecolor{YELLOW}{cmyk}{0,0,1,0}
 }

\makeatother

\usepackage{babel}
\begin{document}

\title{Separating pairing from quantum phase coherence dynamics above the
superconducting transition by femtosecond spectroscopy.}

\author{I.Madan$^{1}$, T.Kurosawa$^{2}$, Y.Toda$^{2}$, M.Oda$^{3}$, T.Mertelj$^{1}$,
P.Kusar$^{1}$, D.Mihailovic$^{1}$}

\affiliation{$^{1}$Jozef Stefan Institute and International Postgraduate School,
Jamova 39, SI-1000 Ljubljana, Slovenia}

\affiliation{$^{2}$Dept. of Applied Physics, University of Hokkaido, Sapporo,
Japan}

\affiliation{$^{3}$Dept. of Physics, University of Hokkaido, Sapporo, Japan}
\begin{abstract}
In classical superconductors an energy gap and phase coherence appear
simultaneously with pairing at the transition to the superconducting
state. In high-temperature superconductors, the possibility that pairing
and phase coherence are distinct and independent processes has led
to intense experimental search of their separate manifestations, but
so far without success. Using femtosecond spectroscopy methods we
now show that it is possible to clearly separate fluctuation dynamics
of the superconducting pairing amplitude from the phase relaxation
above the critical transition temperature. Empirically establishing
a close correspondence between the superfluid density measured by
THz spectroscopy and superconducting optical pump-probe response over
a wide region of temperature, we find that in differently doped $Bi_{2}Sr_{2}CaCu_{2}O_{8+\delta}$ 
crystals the pairing gap amplitude monotonically extends well beyond
$T_{c}$, while the phase coherence shows a pronounced power-law divergence
as $T\rightarrow T_{c}$, thus showing for the first time that phase
coherence and gap formation are distinct processes which occur on
different timescales.
\end{abstract}
\maketitle
Anomalous normal state behavior above the critical temperature appears
to be a hallmark of unconventional superconductivity and is present
in many different classes of materials. A pseudogap state has been
suggested to be associated with a wide range of possible phenomena
preceding the onset of macroscopic phase coherence at the superconducting
critical transition temperature at $T_{c}$: pre-formed pairs \citep{Alexandrov2001,Alexandrov1981,Alexandrov2011a,Mihailovic2002,Kresin2011,Ovchinnikov2002,Geshkenbein1997,Alexandrov1993,Solovev2009},
a spin-gap \citep{EmeryPRB1997}, the formation of a Bose metal \citep{Phillips2003},
a Fermi or Bose glass, or a state composed of {}``dirty bosons''
\citep{Das_PRB1999,Das_PRB2001,Vojta2003}, and more recently a charge-density-wave
state \citep{Torchinsky2013,Sugai2006}.

However, apart from the pseudogap (PG) response below the temperature
designated as $T^{*}$, the response attributed to {}``superconducting
fluctuations'' above $T_{c}$ has been observed in a number of experiments
\citep{Silva2001,Truccato2006,Orenstein2006,Junod2000,Tallon2011,Kondo2010,Wang2006,Rullier-Albenque2006,Pourret2006,Li2010,Mihailovic1987}.
The temperature region $T_{c}<T<T_{onset}$ where such fluctuations
are observable is significantly wider than in conventional superconductors,
but smaller than $T^{*}$. The open and obvious question is whether
the pseudogap, or the superconducting fluctuations can be attributed
to pairing.

The problem in separating the response due to superconducting fluctuations
from the PG is that so far, inevitably, one has had to make extrapolations,
or assumptions about the response functions underlying temperature
dependences and line shapes in transport\citep{Silva2001,Truccato2006,Ghosh1999,Balestrino2001,Bhatia1994},
magnetic susceptibility\citep{Li2010,Wang2005}, specific heat\citep{Junod2000,Tallon2011}
or photoemission (ARPES)\citep{Kondo2010}, which may at best introduce
inaccuracies in the temperature scales, and at worst lead to erroneous
conclusions. Alternatively one can suppress superconductivity by high
magnetic fields up to 60 T \citep{Rullier-Albenque2011}, although
there exists a risk of inducing new states by such a high field \citep{Chen2002}.
Thus, so far it has not been possible to satisfactorily characterize
superconducting fluctuations and discriminate between fluctuations
of the amplitude $\delta\psi$ (related to the pairing gap) and phase
$\delta\theta$ of the complex order parameter $\Psi=\psi e^{i\theta}$.

\begin{figure*}[t]
\includegraphics[width=1\textwidth]{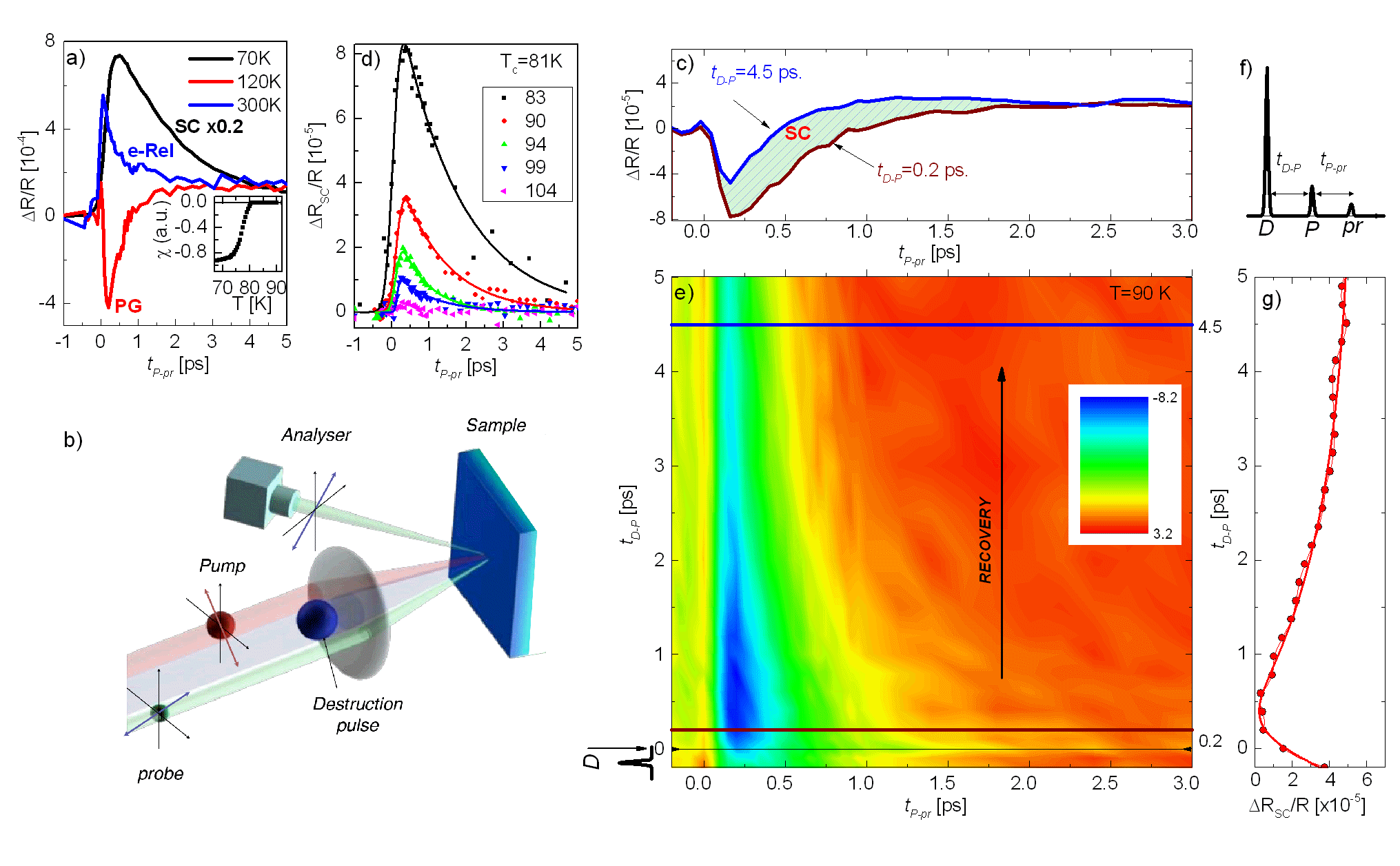}

\caption{\textbf{Description of the 3 pulse method and pump-probe signals.
}Underdoped sample data: a) Transient reflectivity signals observed
at different temperatures in Bi-2212 - SC signal, the PG signal and
energy relaxation of hot carriers (Inset: susceptibility curve shows
$T_{c}=81$ K for the underdoped sample). b) A schematic representation
of the 3-pulse experiment. Destruction (D) pulse destroys the superconductivity,
Pump-probe sequence probes the recovery of the quasiparticle response.
Colors are schematic. c) Two typical 3-pulse pump-probe traces at
$t_{D-P}$ 0.2 and 4.5 ps show signal with suppressed and recovered
superconducting component respectively. d) The QP recombination at
different $T$ above $T_{c}$. e) A typical result of three pulse
experiments at 90K shows suppression of the SC component after the
D pulse arrival and gradual recovery with $t_{D-P}$. Readings along
the blue and dark-red line are shown in Fig. \ref{Fig:Signal_traces}b).
The color represents amplitude of the reflectivity change. The values
of the color bars indicate $\Delta R/R\times10^{-5}$. f) Pulse sequence
and delays notation in 3-pulse experiment. g) Recovery of the superconducting
component amplitude with $t_{D-P}$. \label{Fig:Signal_traces}}
\end{figure*}

In pump-probe experiments three relaxation components shown in Fig.
\ref{Fig:Signal_traces} a) are typically observed: 1) the quasiparticle
(QP) recombination in the SC state, 2) pseudogap state response below
$T*$ and 3) energy relaxation of hot electrons. The QP dynamics has
been shown to be described very well by the Rothwarf-Taylor (R-T)
model \citep{RT_Kaindl2005,RT_Kabanov2005}, and the response related
to the presence of non-equilibrium QPs is thus unambiguous. Importantly,
the presence of the QP response is directly related to the presence
of a pairing gap for QP excitations.

Recent experiments have already proved the coexistence of the pseudogap
excitations with superconductivity below $T_{c}$ over the entire
range of phase diagram \citep{Liu2008,Nair2010}. However, superconducting
fluctuation dynamics above $T_{c}$ have not been investigated till
now. In this paper we present measurements by a 3-pulse technique
which allows us to single out the response of superconducting gap
fluctuations, distinct from the PG. Selective destruction of the superconducting
state by a femtosecond laser pulse \citep{Kusar_PRL2008} allows us
to discriminate pseudogap excitations from superconducting fluctuation
seen in transient reflectivity signals, thus avoiding the necessity
of making extrapolations or assumptions in separating the different
contributions. We then compare these data with a.c. conductivity (THz)
measurements \citep{CorsonORENST1999} and establish proportionality
of the amplitude of superconducting component in pump-probe experiment
to the bare phase stiffness $\rho_{0}$, measured by THz experiments.
This is directly proportional to the bare pair density $n_{s}$ \citep{CorsonORENST1999,Wang2006},
which in turn coincides with the superfluid density when the latter
is measured on a timescale on which changes of the order parameter
due to either de-pairing or movement of the vortices can be neglected.
Comparison of the critical behavior of the amplitude and phase correlation
times above $T_{c}$ leads us to the conclusion that two quantities
arise from different microscopic processes.

We perform measurements on under- (UD), near optimally-(OP) and over-(OD)
doped Bi2212 with $T_{c}$s of 81, 85 and 80 K respectively. In the
discussion we focus on the underdoped sample, and discuss comparisons
with the optimally and overdoped samples, where applicable.

\section*{Results}

\textbf{Measurements of pairing amplitude above $T_{c}$}. To separate
the SC component from the PG component we use a 3 pulse technique
described in Refs. \citep{Kusar_Submited2013,Yusupov_NatPhys2010,Kusar_PRB2011,Mertelj_PRL2013},
and schematically represented in Fig. \ref{Fig:Signal_traces}b).
A pulse train of 800 nm 50 fs pulses produced by a 250 kHz regenerative
amplifier is divided into three beams with variable delays. First
a relatively strong {}``destruction'' (D) pulse, with fluence just
above the superconducting state photo-destruction threshold $F_{th}^{SC}=13$
$\mu J/cm^{2}$\citep{Toda_PRB2011}, destroys the superconducting
condensate\citep{Stojchevska2011}. The ensuing recovery of the signal
is measured by means of the 2 pulse Pump-probe (P-pr) response at
a variable delay $t_{D-P}$ between D and P pulses. The pseudogap
state remains unaffected as long as the excitation fluence is well
below the pseudogap destruction fluence which is measured to be at
$F_{th}^{PG}=32$ $\mu J/cm^{2}$. Measurements at higher temperatures
(at 120 and 140 K) where no fluctuations are present confirm that
the D pulse has no effect on the PG response at the selected fluence.

\begin{figure}[H]
\includegraphics[width=1\columnwidth]{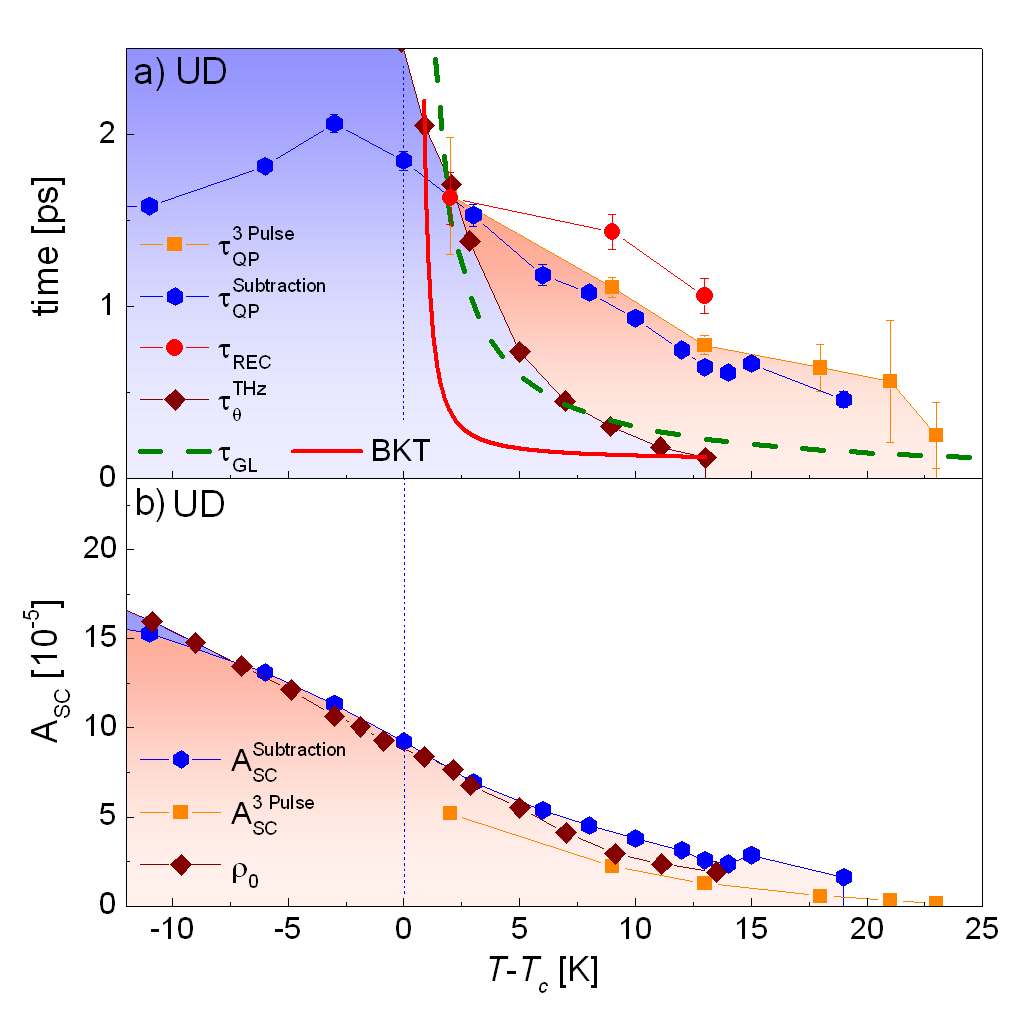}

\caption{\label{Fig.The-relaxation-times} \textbf{Comparison of pairing amplitude
and phase coherence dynamics. }a) The recovery time of the optical
superconducting signal ($\tau_{rec}$), the QP recombination time
$\tau_{QP}^{3Pulse}$ measured by the three pulse technique and the
QP recombination time $\tau_{QP}^{Subtraction}$ from two pulse Pump-probe
pulse measurements obtained by subtraction of the PG signal. A fit
to the data using a BKT model (eq. 3 of ref. \citep{Orenstein2006})
is shown by the solid red line. The dashed line shows the fluctuation
lifetime $\tau_{GL}$ given by the TDGL theory. The phase correlation
time $\tau_{\theta}^{THz}$ obtained from the THz conductivity measurements
\citep{CorsonORENST1999} is also shown for comparison. b) The amplitude
of the SC signal measured by the three pulse technique ($A_{SC}^{3Pulse}$)
and the two pulse measurements with the PG signal subtracted ($A_{SC}^{Subtraction}$).
The bare phase stiffness $\rho_{0}$ \citep{CorsonORENST1999} (normalized
at $T_{c}$) shows remarkable agreement with the optical response. }
\end{figure}

A typical result of the 3 pulse experiment is presented in Fig. \ref{Fig:Signal_traces}e).
In the absence of the D pulse the signal consists of a positive SC
and a negative PG component. After the arrival of the D pulse we see
a disappearance of the SC part and only the PG component is present
(dark-red line on Fig. \ref{Fig:Signal_traces}c)). With increasing
$t_{DP}$ the superconducting response gradually re-emerges (blue
line on Fig. \ref{Fig:Signal_traces}c)).  As most of the condensate
in the probe volume is {}``destroyed'' by the D pulse we can extract
the superconducting component by subtracting the signal remanent after
the destruction (measured 200 fs after the D pulse) from the signal
obtained in the absence of the D pulse. Such an extracted superconducting
component is plotted in Fig. \ref{Fig:Signal_traces}d). The signal
is detectable up to $T_{onset}=104$ K, which is 0.28 $T_{c}$ above
$T_{c}$ but much lower than $T^{*}\approx2.5$ $T_{c}$ . In Fig.
\ref{Fig:Signal_traces}g) we show the recovery of the amplitude of
the superconducting component $A_{sc}$ as a function of the time
delay $t_{D-P}$ after the D pulse for 90 K. The temperature dependence
of the amplitude of the SC component measured by 3-pulse technique
$A_{sc}^{3pulse}$ is shown in Fig. \ref{Fig.The-relaxation-times}b).

For comparison, standard pump-probe measurements need to separate
the SC relaxation from the PG relaxation by subtraction of the high
temperature response extrapolated into the superconducting region.
This approach suffers from the same uncertainties as other techniques
such as conductivity, heat capacity, diamagnetism and ARPES. The actual
$T$-dependence of the PG response is ca vary with doping, pump energy
and probe wavelength \citep{Toda_PRB2011,Coslovich2013}. But, in
Fig. \ref{Fig.The-relaxation-times}b) we show that the subtraction
procedure - with the use of a model \citep{Toda_PRB2011} - gives
results in agreement with the direct 3 pulse measurements. The remaining
discrepancies in the amplitude can be explained by an incomplete destruction
of fluctuating superconducting state in the 3 pulse experiment and
errors in the PG subtraction.

In Fig. \ref{Fig.The-relaxation-times}a) we see that the QP relaxation
time $\tau_{QP}^{3Pulse}$ obtained by fitting an exponential function
to the data Fig. \ref{Fig:Signal_traces}d) decreases rather gradually
with increasing $T$ above $T_{c}$ and nearly coincide with the QP
relaxation time obtained by the pseudogap subtraction procedure $\tau_{QP}^{Subtraction}$.
The recovery time $\tau_{rec}$ obtained from exponential fits to
the recovery of the SC response above $T_{c}$ (Fig. \ref{Fig:Signal_traces}
e) shows a similar $T$-dependence. The experiments thus show that
the recovery of the SC gap and the QP relaxation show very similar
dynamics above $T_{c}$.

\textbf{Comparison of optical and a.c. conductivity measurements.}
We now compare these data with THz measurements of the order parameter
correlation time and bare phase stiffness \citep{Orenstein2006,CorsonORENST1999}.
The agreement between $\rho_{0}$ and $A_{sc}$ shown in Fig. \ref{Fig.The-relaxation-times}
b) is seen to be remarkably good over the entire range of measurements
$0.8\, T_{c}<T<1.3\, T_{c}$. This agreement is important because,
taking into account $n_{s}\sim|\Psi|{}^{2}$, it validates the approximation
that the pump-induced changes in the reflectivity or dielectric constant
$\epsilon$ for small $n_{s}$ are related to the order parameter
$\Psi$ as $\delta R\sim\delta\varepsilon\sim|\Psi|{}^{2}$.

In contrast to $A_{SC}$ and $\rho_{0}$, remarkable differences are
seen in the temperature dependences of the \emph{characteristic lifetimes}
shown in Fig. \ref{Fig.The-relaxation-times}a) obtained by optical
techniques and THz conductivity measurements. The phase correlation
time $\tau_{\theta}^{THz}$ determined from the THz conductivity \citep{Orenstein2006}
dies out very rapidly with increasing temperature, while the $T$-dependence
of the amplitude relaxation ($\tau_{QP}^{3Pulse}$, $\tau_{QP}^{Subtraction}$
or $\tau_{REC}$) is much more gradual.

\begin{figure*}[tp]
\includegraphics[width=1\textwidth]{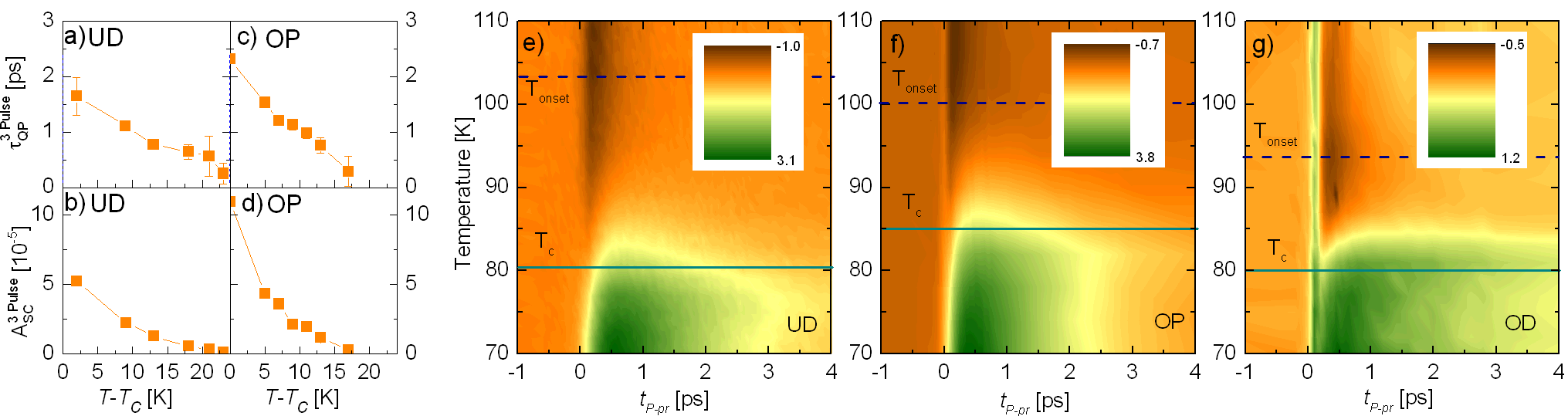}

\caption{\textbf{Doping dependence.} Comparison of the QP recombination time
$\tau_{QP}^{3Pulse}$ a) and c) and the amplitude of the SC signal
b) and d) measured by the three pulse technique for under-(UD) and
optimally(OP) doped samples, respectively. e) - g) T-dependence of
2-pulse response for under-(UD), optimally(OP) and overdoped(OD) samples.
Blue dashed and cyan solid lines marks $T_{onset}$ and $T_{c}$ respectively.
$T_{onset}$ shows gradual decrease with doping. The values of the
color bars indicate $\Delta R/R\times10^{-4}$.\label{fig: Doping}}
\end{figure*}

Measurements on an optimally doped sample (Fig. \ref{fig: Doping}c),
d)) show qualitatively the same results with $T_{onset}\sim$102 K,
which is 17 K above $T_{c}$, and slightly faster decrease of both
amplitude and QP relaxation time with temperature. For the overdoped
sample, $F_{th}^{PG}$ becomes comparable to $F_{th}^{SC}$, so the
superconducting component cannot be significantly suppressed without
affecting the pseudogap. Nevertheless the superconducting component
is clearly observable in the 2-pulse response up to $T_{onset}\sim93$
K, i.e. 13 K above $T_{c}$. Comparison of 2-pulse data for different
doping levels is shown in Fig. \ref{fig: Doping} e)-g), showing qualitatively
similar behavior of the SC amplitude above $T_{c}$.

\section*{Discussion}

The co-existence and distinct dynamics of the PG and SC excitations
above $T_{c}$ highlights the highly unconventional nature of these
states. A possible explanation for the coexistence of the SC and
PG excitations is that the SC and PG quasiparticles which are giving
rise to the observed processes are associated with relaxation at different
regions on the Fermi surface. Recent Raman and cellular dynamical
mean-field studies \citep{Sakai2013} have suggested that the PG may
originate from the states inside the Mott gap, which are characterized
by s-wave symmetry and very weak dispersion. Such a localized nature
of the PG state excitations is consistent with previous assignments
made on the basis of pump-probe experiments \citep{Kabanov_PRB1999,Toda_PRB2011}.
In contrast, the superconducting gap fluctuations have predominantly
$d$-wave symmetry \citep{Toda2013} and are more delocalized. This
would explain the simultaneous presence of the SC fluctuations and
PG components in pump-probe experiments.

Perhaps the most widely discussed model in the context of distinct
pairing and phase coherence phenomena is the Berezinskii-Kosterlitz-Thouless
(BKT) transition \citep{Berezinskii1971,Berezinskii1972,Kost1973}
by which decreasing temperature through $T_{onset}$ and approaching
$T_{c}$ causes freely moving thermally activated vortices and anti-vortices
to form pairs, thus allowing the condensate to acquire long range
phase coherence in an infinite-order phase transition sharply at $T_{c}$.
The bare pair density is finite to much higher temperatures (up to
$T_{onset}$), where pairing is caused by a different mechanism, and
the pseudogap is considered to be an unrelated phenomenon\citep{Emery1995,Emery1997,Emery1998,Fisher1991}.
The effect is \foreignlanguage{british}{characterised} in terms of
a phase stiffness $\rho_{s}$, a quantity which characterizes the
destruction of phase coherence by thermal fluctuations at a temperature
$T_{c}=T_{BKT}=\pi\rho_{s}/8$. It is defined by the free energy cost
of non-uniformity of the spatially varying order parameter $\Psi$
\citep{Orenstein2006}. In cuprates $\rho_{s}$ is small due to reduced
dimensionality and the low carrier density. Within this approach a.c.
(THz) conductivity\citep{CorsonORENST1999,Orenstein2006}, heat capacity
\citep{Junod2000,Tallon2011}, diamagnetism \citep{Li2010,Wang2005},
ARPES \citep{Kondo2010} and Nernst effect \citep{Wang2006,Rullier-Albenque2006,Pourret2006,Mukerjee2004}
measurements were interpreted.

The BiSCO family is the most two-dimensional of the cuprate materials,
so here the BKT mechanism for describing the order above $T_{c}$
would be expected to be most applicable \citep{CorsonORENST1999}.
However, the data in Fig. \ref{Fig.The-relaxation-times}a) show that
the drop in $\tau_{\theta}$ is not nearly as abrupt as the BKT model
predicts. One possibility is that this broadening arises from chemical
inhomogeneity of the sample, but the absence of a peak in the heat
capacity at $T_{onset}$ \citep{Junod2000} also appears to exclude
the possibility of a pure BKT transition, and implies amplitude fluctuations
might be present between $T_{c}$ and $T_{onset}$ as well \citep{Tallon2011}.
Thus additional mechanisms beyond BKT may also be present which would
broaden the phase coherence transition, such as inter-layer phase
fluctuations. In this case the observed $T$-dependence would reflect
the interlayer de-coherence \citep{CorsonORENST1999,Orenstein2006}.

In more traditional approaches using time-dependent Ginzburg-Landau
(TDGL) theory\citep{Larkinlate2005}, thermal fluctuations are small
for temperatures higher than $\sim2$ K above $T_{c}$ \citep{VanderBeek2000},
but can give an observable contribution to the conductivity in this
temperature range \citep{Silva2001,Truccato2006,Ghosh1999,Balestrino2001,Bhatia1994}.
Relaxation within TDGL theory has a \textquotedblleft{}longitudinal\textquotedblright{}
relaxation time $\tau_{\Delta}$, which corresponds to the relaxation
of the magnitude of the SC order parameter, and a \textquotedbl{}transverse\textquotedbl{}
relaxation time $\tau_{\theta}$ which corresponds to the relaxation
of its phase. They are related to each other in magnitude, but have
the same critical temperature dependence near $T_{c}$, namely $\tau_{GL}\sim1/(T-T_{c})$
\citep{PhysRevLett.36.429}. Perhaps unexpectedly, the temperature
dependence of $\tau_{\theta}^{THz}$ nearly coincides with the behavior
predicted for $\tau_{\theta}$ by time-dependent Ginzburg Landau theory
for Gaussian fluctuations. Fig. \ref{Fig.The-relaxation-times}a)
suggests that phase coherence within this system is established in
a narrow, $\sim5$ K temperature interval. However, the distinctly
different critical behavior of the pair amplitude dynamics speaks
in favor of unconventional models of superconductivity in which pairing
and phase coherence occur independently, by different mechanisms.
The implication is that the observed pairing amplitude which extends
to more than 25 K above $T_{c}$ reflects the response of an inhomogeneous
ensemble of gapped patches which are not mutually phase coherent.
The weak temperature dependence of the amplitude cannot be described
either by TDGL or BKT models.

Beyond the BKT vortex and TDGL scenario, other phase-locking scenarios,
such as Bose-Einstein condensation of bipolarons\citep{Alexandrov1981,Alexandrov2011,Alexandrov2011a}
and phase-coherence percolation\citep{Kresin2011,Mihailovic2002}
may also be consistent with the observed \foreignlanguage{british}{behaviour}.
In both of these cases pairing and phase coherence are also distinct
processes. The former comes from the condensation of pairs at $T_{c}$
as pre-formed pair kinetic energy is reduced, while percolation dynamics
is associated with the time dynamics of Josephson tunneling between
fluctuating pairs or superconducting patches. The percolation timescale
$\tau_{J}$ is given by the Josephson energy $E_{J}=I_{c}\phi_{0}/2\pi$,
where $I_{c}$ is the critical current and $\phi_{0}$ is the flux
quantum. In cuprates, $\tau_{J}=\hbar/E_{J}\simeq300$ fs, which is
compatible with the dynamics of phase shown in Fig. \ref{Fig.The-relaxation-times}a).

A picture highlighted by Fig. \ref{Fig.The-relaxation-times} thus
emerges in these materials where the relaxation of the phase $\theta$
is faster than relaxation of the amplitude $\psi$ of the complex
order parameter $\Psi=\psi e^{i\theta}$, the dynamics of $\psi$
and $\theta$ being governed by microscopically different processes.
It is worth remarking here that the opposite situation is found in
charge density wave dynamics, where the phase relaxation is slow compared
to the amplitude relaxation, and the dynamics can be described by
TDGL equations for the amplitude $\psi$, neglecting phase relaxation
$\theta$ \citep{Yusupov_NatPhys2010}.

\section*{Methods}

\textbf{Samples.} The samples used in this work were under-, near
optimally- and over- doped Bi2212 with $T_{c}$s of 81, 85 and 80
K respectively. Samples were grown by the traveling solvent floating
zone method. Critical temperatures were obtained from susceptibility
measurements (e.g. inset in Fig. \ref{Fig:Signal_traces}a) for the
underdoped sample).

\section*{Author contributions}

T.K., Y.T. and M.O. has grown the samples and done magnetic characterisation,
I.M. did optical measurements. I.M. and P.K. performed data analysis.
I.M., Y.T, T.M. and D.M. interpreted the data. I.M and D.M. wrote
the manuscript.

\section*{Additional information}

\textbf{Competing financial interests:} The authors declare no competing
financial interests.
\end{document}